\documentclass[10pt,prd,twocolumn,preprint,nofootinbib]{revtex4}

\usepackage{epsfig}
\usepackage{amsmath,amsfonts,amssymb}
\usepackage{t1enc}
\usepackage{verbatim}
\usepackage{float}
\usepackage{morefloats}
\usepackage{color}

\providecommand{\openone}{\leavevmode\hbox{\small1\kern-3.8pt\normalsize1}}

\usepackage{booktabs}
\usepackage[Q=yes,pverb-linebreak=no]{examplep}

\newcommand{\be}{\begin{equation}}
\newcommand{\ee}{\end{equation}}

\begin{document}

\title{\boldmath Reconstruction of top quark pair dilepton decays in electron-positron collisions \unboldmath}

\author{
Helenka~Casler$^{1,2}$,
Matthew~Manganel$^{3}$,
Miguel~C.~N.~Fiolhais$^{4,5}$,
Andrea~Ferroglia$^{1,6}$,
Ant\'onio Onofre$^{7}$
\\[3mm]
{\footnotesize {\it 
$^1$The Graduate School and University Center, The City University of New York, 365 Fifth Avenue, New York, NY 10016  USA \\
$^2$Department of Earth and Physical Sciences, York College, City University of New York,\\ 94 - 20 Guy R. Brewer Blvd, Jamaica, NY 11451, USA \\
$^3$ Courant Institute of Mathematical Sciences, New York University, 251 Mercer Street, New York, NY 10012, USA \\
$^4$ Science Department, Borough of Manhattan Community College, City University of New York, \\ 199 Chambers St, New York, NY 10007, USA \\
$^5$ LIP, Departamento de F\'{\i}sica, Universidade de Coimbra, 3004-516 Coimbra, Portugal\\
$^6$Physics Department, New York City College of Technology, The City University of New York, 300 Jay Street, Brooklyn, NY 11201, USA \\
$^7$ Departamento de F\'{\i}sica, Universidade do Minho, 4710-057 Braga, Portugal\\
}}
}

\begin{abstract}
A new algorithm is presented to perform the full kinematic reconstruction of top quark pair  events produced at future electron-positron colliders in the case of dilepton decays of the $W$ bosons to electrons or muons. The momentum components of the undetected neutrino and anti-neutrino in the event are reconstructed by employing several kinematic conditions comprising a non-linear system of six equations. This system is solved numerically using two independent methods and the selection of the best candidate real solution for each event is determined by a likelihood discriminant. Results are presented for several reconstructed kinematic properties of the $W^\pm$ bosons, top (and anti-top) quarks using generator level information produced at leading order.
\end{abstract}

\maketitle

\section{Introduction}
\label{sec:intro}
While the LHC is producing extremely precise measurements of top quark and Higgs boson properties, the planning for the next high energy particle accelerator is already underway. The objective of the planning and designing work is to develop a machine able to investigate in depth beyond the Standard Model (SM) physics scenarios. Despite the fact that there is no decision yet to determine which accelerator will be built, or its location, there is a consensus in the scientific community that the results from the LHC will have to be complemented by an accelerator that can measure observables with greater precision by producing high energy collisions between electrons and positrons. At present, the most likely candidates are the International Linear Collider (ILC)~\cite{Behnke:2013xla}, the Compact Linear Collider (CLIC)~\cite{Linssen:2012hp} and the  Future Circular Collider (FCC)~\cite{Benedikt:2653669,Ellis:2015sca}. The ILC is expected to operate at center of mass energy of 500~GeV with a possible upgrade to 1~TeV. CLIC is designed to collide electrons and positrons at a nominal energy of 3~TeV, {while FCC is expected to operate with a center of mass energy in the range of 90--350~GeV}.

Future electron-positron accelerators comprise several advantages which allow for high-precision phenomenological studies in Higgs boson processes, studies of top-quark pair production and searches for new particles. One of the advantages provided by an electron-positron machine is the ability to operate  within a wide range of center-of-mass energies. An electron-positron collider also makes it possible to collide electron and positron beams with high spin polarization. This feature opens the window to several new observables that could not be measured using hadron colliders~\cite{AguilarSaavedra:2012vh,Fiolhaisconf}. In addition, high energy collisions in electron-positron accelerators are less complex when compared to proton-proton collisions. As a result, particle detectors in electron-positron colliders have higher intrinsic resolution than those at machines colliding protons.

The reconstruction of neutrinos is one of the main experimental challenges in high energy physics experiments as they do not interact with the active material of the particle detector. As such, the momenta of the neutrinos are normally associated with the missing momentum in the event resulting from an high-energy collision. This leads to a straightforward reconstruction if only a single neutrino (or anti-neutrino) is produced in the physics process of interest. The reconstruction becomes highly non-trivial if two or more neutrinos are produced. This particular problem has been successfully addressed at the LHC in top-quark pair production and Higgs boson production in association with a top-quark pair. The full kinematic reconstruction of the four-momenta of the undetected neutrinos was performed by imposing energy-momentum conservation, and using mass constraints on the top quarks and $W^\pm$~bosons ~\cite{Aad:2012ky,Santos:2015dja,AmorDosSantos:2017ayi,Azevedo:2017qiz}. 

In this paper, a similar strategy is implemented in order to develop an algorithm to reconstruct the momentum of the neutrino (and anti-neutrino) resulting from the dilepton decay of a top-quark pair ($t\bar{t}$) produced in an electron-positron collision. By using the momenta of all detected final state particles resulting from the top and anti-top quark decays, a system of six kinematic equations is implemented in order to determine the unknown momenta of the neutrino and anti-neutrino. The system is solved by using two different numerical methods applied to one-million event samples generated with \verb!MadGraph5_aMC@NLO! at leading order~\cite{Alwall:2014hca}. This particular non-linear system of equations leads to multiple possible complex and real solutions for the neutrino's momenta. Consequently, an extensive statistical study is performed to determine which physical variables are the best decision-making indicators to choose the best candidate real solution by means of a likelihood method. Results are presented by comparing several reconstructed kinematic properties such as the transverse momentum and masses of the  \mbox{(anti-) top} quarks and $W^\pm$ bosons with the generator level information provided by \verb!MadGraph5_aMC@NLO!. This study is performed at parton level for events which have tree-level kinematics. The effects of background events, beam resolution, radiation, detector simulation and selection cuts are beyond the scope of the manuscript, and are therefore deferred to a future study. The reconstruction code packages are made available in public repositories~\cite{link_to_package1,link_to_package2}.

\section{Kinematic Equations}

The reconstruction of $t\bar t$ dilepton events assumes the experimental detection of two $b$-tagged jets and two opposite charged leptons together with the measurement of missing energy associated with the neutrino and anti-neutrino. The reconstruction procedure is not expected to be applied to events with additional objects, such as a hard photon or an additional jet. Neutrinos are not detected as they escape without interacting with the active material of the detector. Consequently, the neutrino momenta can be associated with the missing energy. Using the mass of the $W^\pm$ bosons as constraints, and assuming the approximation that all final state particles are massless, the three-momenta of the neutrino and anti-neutrino can be determined from six kinematic equations. To begin with, three linear equations can be written as,
\begin{eqnarray}
\ensuremath{p_{i}^{\nu} + p_{i}^{\bar{\nu}} &=& p_{i}^{\textrm{miss}}}\, , 
\label{eq:linear}
\end{eqnarray}
where $p_{i}^{\textrm{miss}}$ represents the components of the missing momentum, and $p_{i}^{\nu}$ and $p_{i}^{\bar{\nu}}$ correspond to the neutrino and anti-neutrino momentum components, respectively. Conservation of center of mass energy at the collision point, $\sqrt{s}$, is employed to obtain a non-linear equation:
\begin{eqnarray}
E^{\ell^-} + E^{\ell^+} + E^{\nu} + E^{\bar{\nu}} + E^{b} + E^{\bar{b}} &=& \sqrt{s} \, ,
\label{eq:energy}
\end{eqnarray}
where $E^{i}$ represents the energy of the particle $i$. Finally, the $W^\pm$ bosons mass are constrained to a fixed value, $m_{{W}} = 80.4$~GeV, leading to two additional quadratic kinematic equations,
\begin{eqnarray}
\ensuremath{( p^{\ell^-} + p^{\bar{\nu}} )^{2} &=& m_{{W}}^2}, \nonumber \\
\ensuremath{( p^{\ell^+} + p^{\nu} )^{2} &=& m_{W}^{2}},
\label{eq:quadratic}
\end{eqnarray}
where $p=(E,\vec{p})$ represents the four-vector of the particles. These six kinematic equations can be used to determine a total of six unknown momentum components for the neutrino and anti-neutrino. However, due to the presence of two quadratic equations (\ref{eq:quadratic}) and one polynomial equation (\ref{eq:energy}) with nested radicals, it is hard to retrieve a set of analytic solutions from this kinematic system of equations. Nonetheless, the analytic elimination of five out of the six unknown variables leads to two implicit non-linear equations for the same unknown quantity, which is taken to be $p_{z}^{{\nu}}$. The physical value of $p_{z}^{{\nu}}$ is a solution of one of these two equations.

The six kinematic equations are significantly different from the ones used at the LHC for the top quark pair production in the same decay channel~\cite{Aad:2012ky,Santos:2015dja,AmorDosSantos:2017ayi,Azevedo:2017qiz}. In particular, at the LHC the total linear momentum of the final state particles caught by the detector and the neutrinos is only zero on the transverse plane, and not along the collision line ($z$-axis). Furthermore, since the proton is not an elementary particle, the center of mass energy of the $t\bar{t}$ system is unknown. {As a  result, the kinematic equations for the reconstruction at the LHC require \mbox{(anti-) top} quark mass constrains, while the reconstruction at the electron-positron collider can be performed without imposing any condition on the kinematics of the \mbox{(anti-) top} quarks.} 

\section{Event Reconstruction \label{sec:reconstruction}}

While it is difficult to retrieve an analytic solution from the system of equations presented in the previous section, it is possible to obtain numerical solutions on a event-by-event basis. As such, two numerical methods were implemented and compared to make sure they provide consistent results. Both methods were applied to a sample of one million $e^-e^+ \rightarrow t\bar{t} \rightarrow W^+ W^- b \bar{b} \rightarrow {\ell}^+ {\ell}^- \nu \bar{\nu} b \bar{b}$ events generated with \verb!MadGraph5_aMC@NLO! at leading order at a center of mass energy of 1~TeV. {This sample was labelled as Sample A. The events in this sample were generated with massless final state particles, with exception of the bottom quarks. The mass of the bottom quarks was set to its on-shell value of 4.7~GeV.}
In the rest of this work, the adjective ``generated'' refers to the list of events and momenta obtained from \verb!MadGraph5_aMC@NLO! as explained above.

The two implicit equations for $p_{z}^{{\nu}}$ were treated as distinct problems to be solved individually. For each event, the equations were put in form,
\begin{equation}
f_i(p_{z}^{{\nu}}) = 0 \, , \,\,\, i=1,2 \, .
\label{eq:f}
\end{equation}

In the first approach, a bisection method was used to find the real solutions of equation~(\ref{eq:f}). The allowed range of $p_{z}^{{\nu}}$ was limited to the real domain of the $f_i(p_{z}^{{\nu}})$.
This range was bisected in search of solutions until the remaining range was less than 1\% the size of the original range. Each one of the functions in equation~(\ref{eq:f}) was found to have one real solution at most.

For the second approach, the functions $f_i(p_{z}^{{\nu}})$ were approximated by interpolating each function with a degree-four polynomial within the allowed range of values for $p_{z}^{{\nu}}$. While analytic solutions to these polynomials exist, it is more efficient to solve them numerically using the method documented in~\cite{numpy}. 
Once the functions are interpolated, the roots of the polynomials should coincide with those of the original functions, provided the interpolation error is small. Therefore, to confirm that the polynomials adequately approximate the functions, the adjusted R-squared coefficient is calculated using a set of fifty test points for each interpolation. Solutions were only taken from interpolants with an R-squared of 0.95 or greater. In one million events, all but a single R-squared coefficient fell above this threshold.  A degree of four was chosen because it is the smallest polynomial degree that fits $f_i(p_{z}^{\nu})$ with this level of precision.

\begin{figure*}
\begin{center}
\includegraphics[width=8.cm]{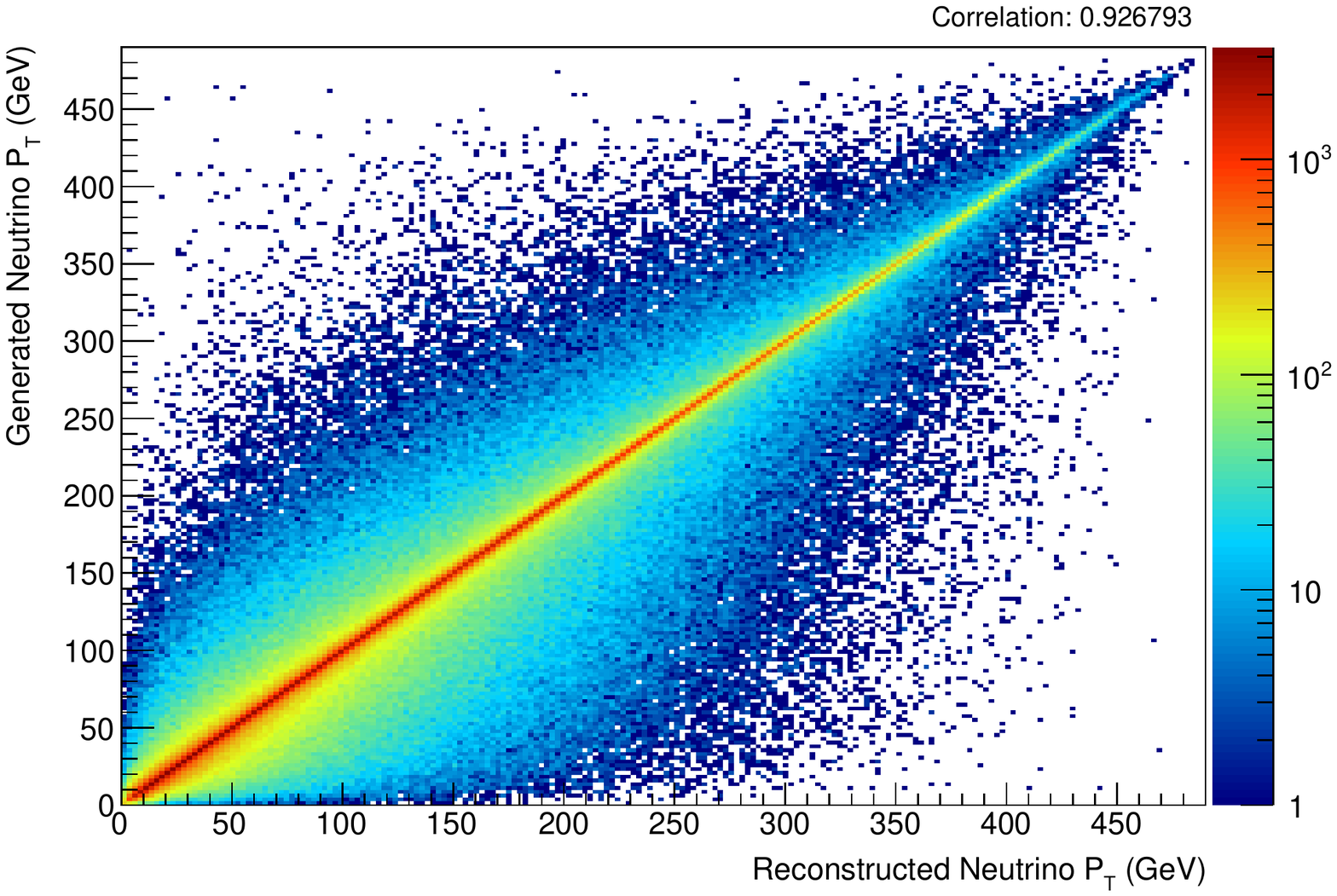}
\includegraphics[width=8.cm]{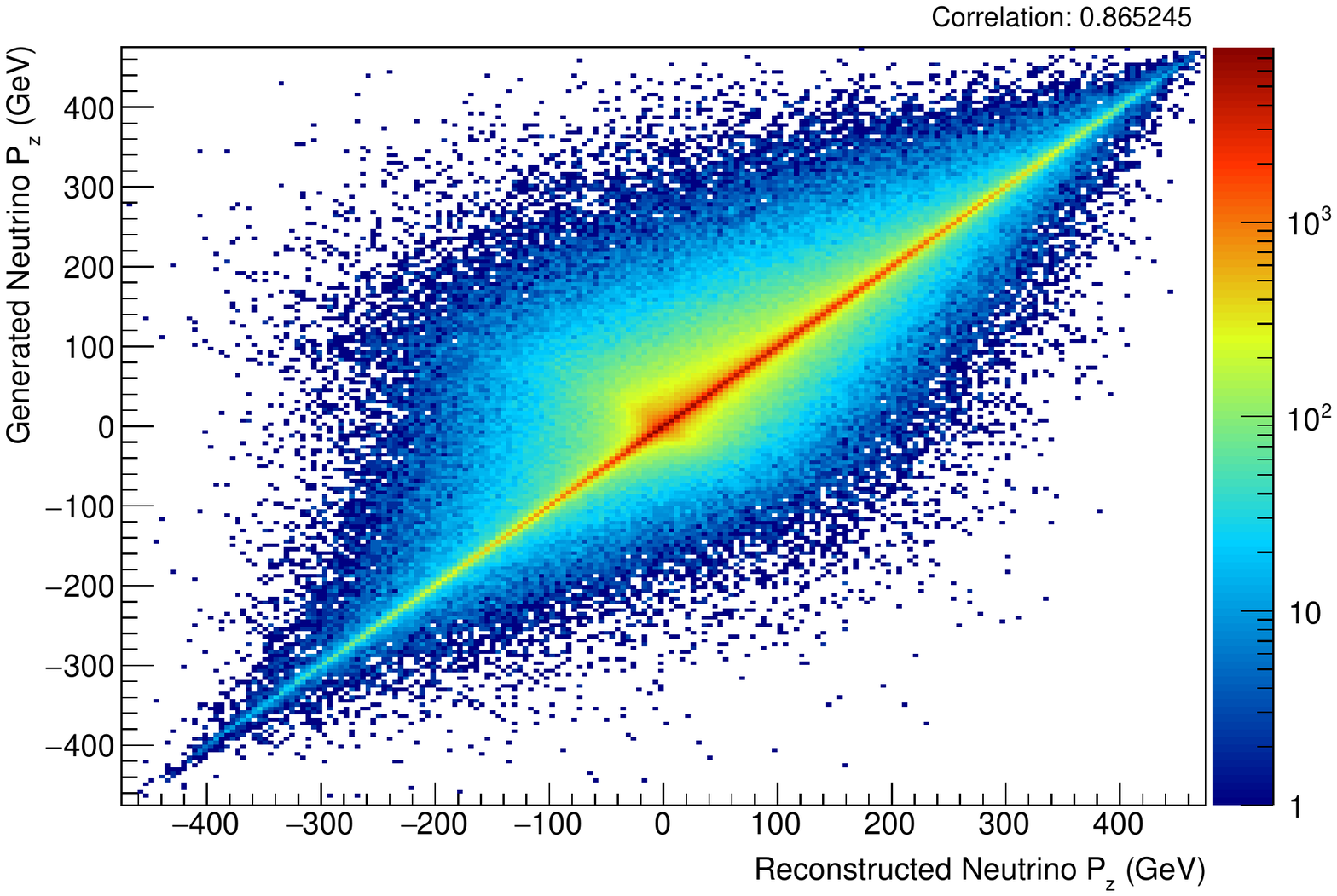} \\
\includegraphics[width=8.cm]{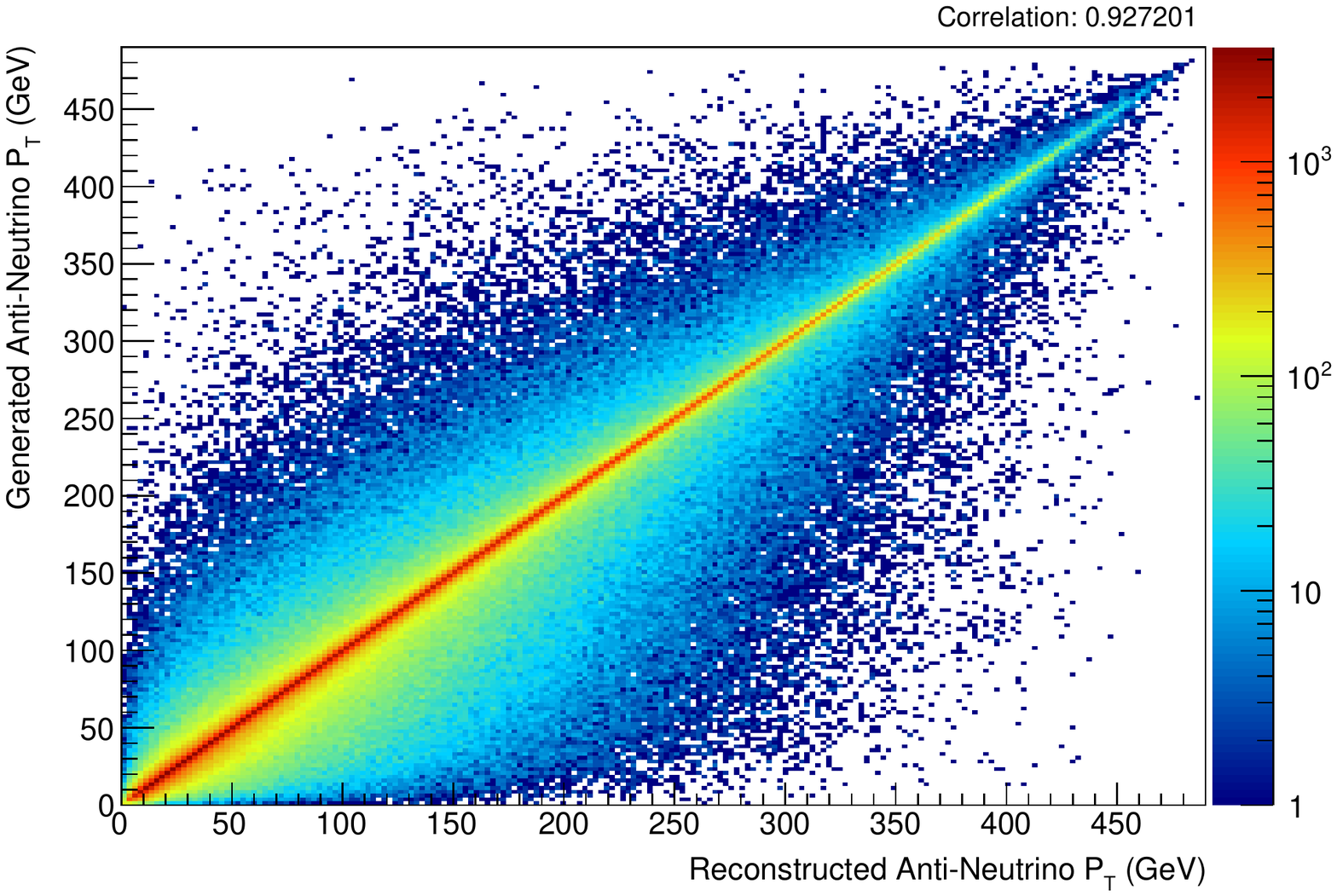}
\includegraphics[width=8.cm]{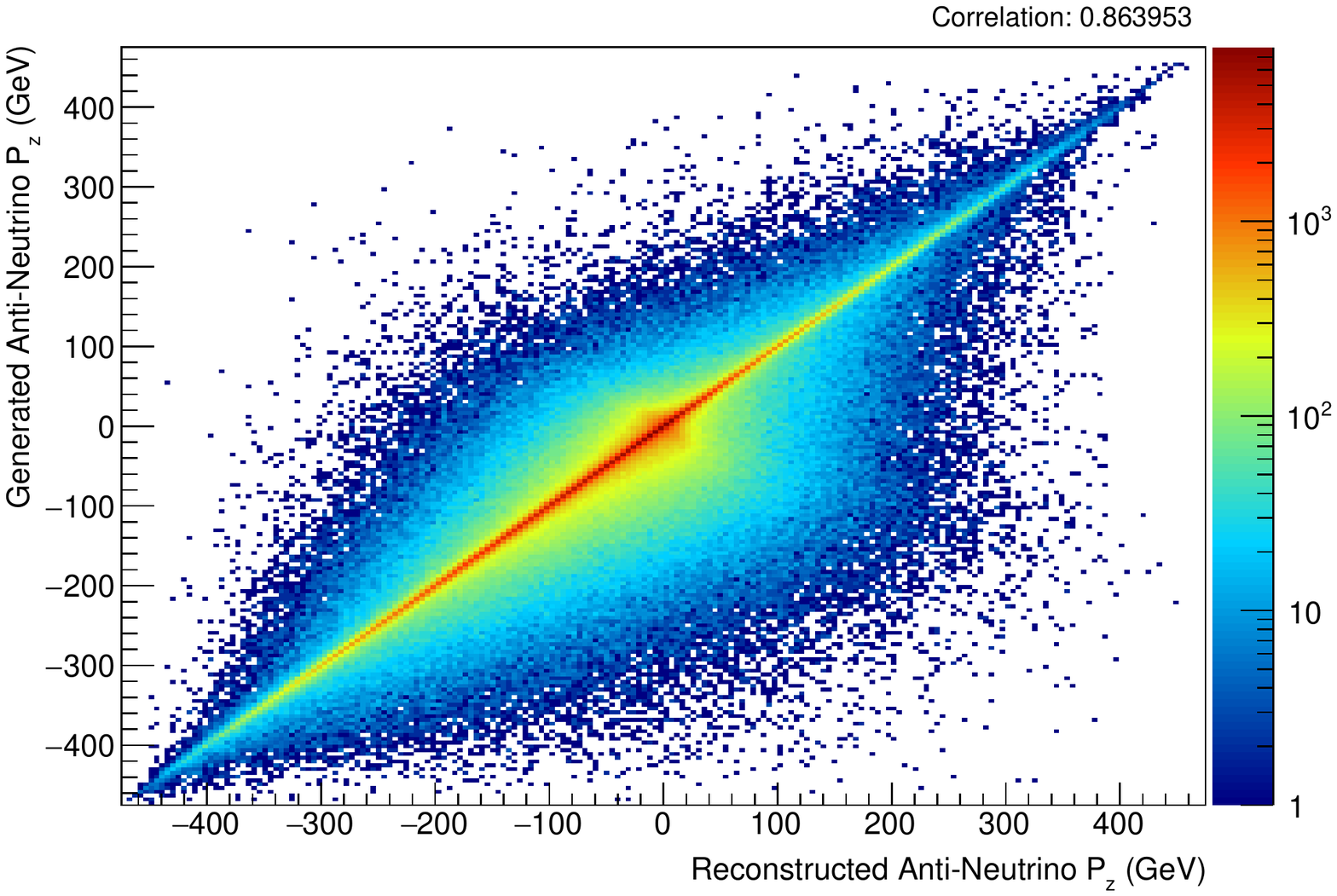}
\caption{Correlation plots between the generated and reconstructed transverse (left) and $z$-axis (right) momentum components of the neutrino (top) and anti-neutrino (bottom). These plots were produced with Sample A.}
\label{fig:random}
\end{center}
\end{figure*}

\begin{figure*}
\begin{center}
\includegraphics[width=8.cm]{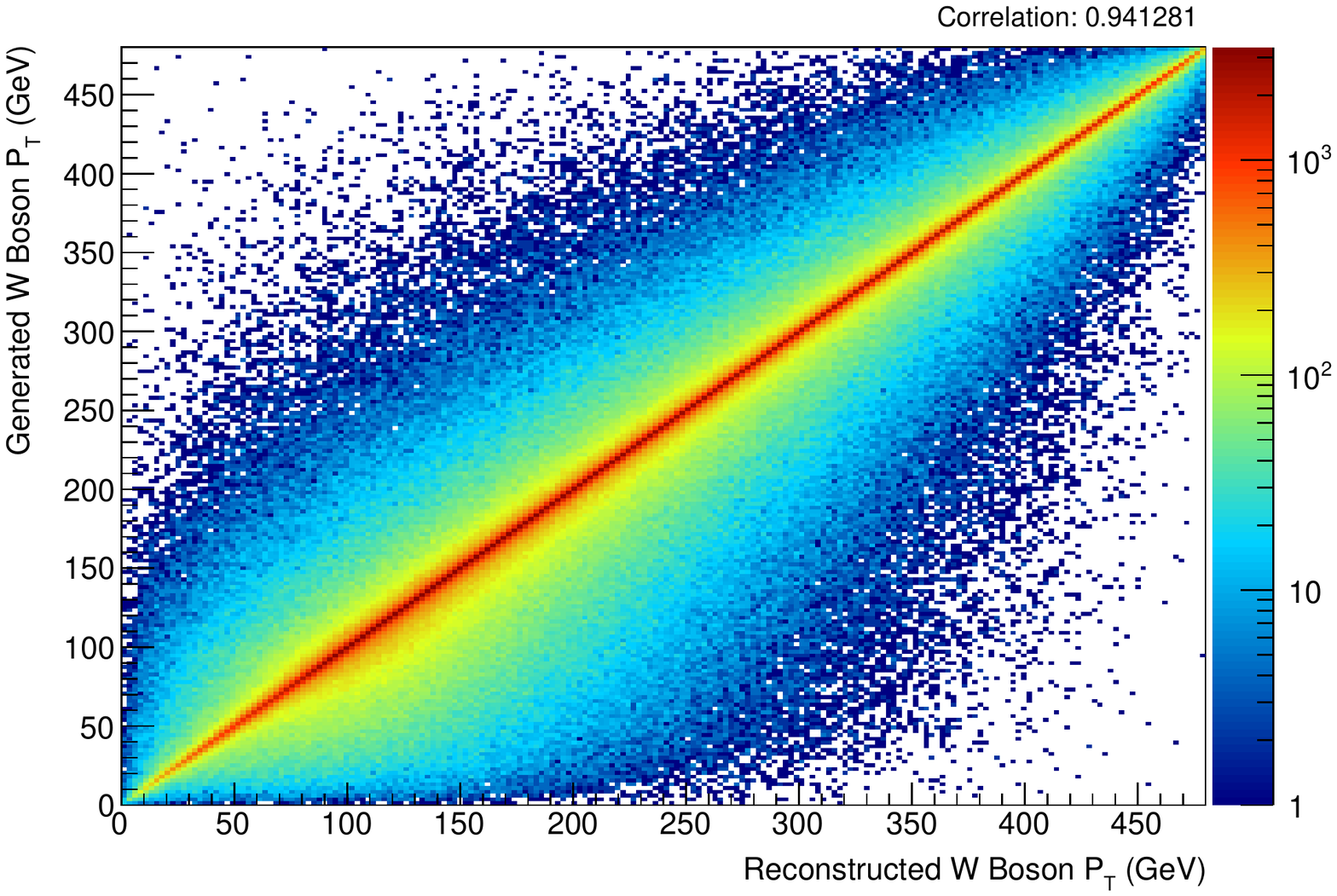}
\includegraphics[width=8.cm]{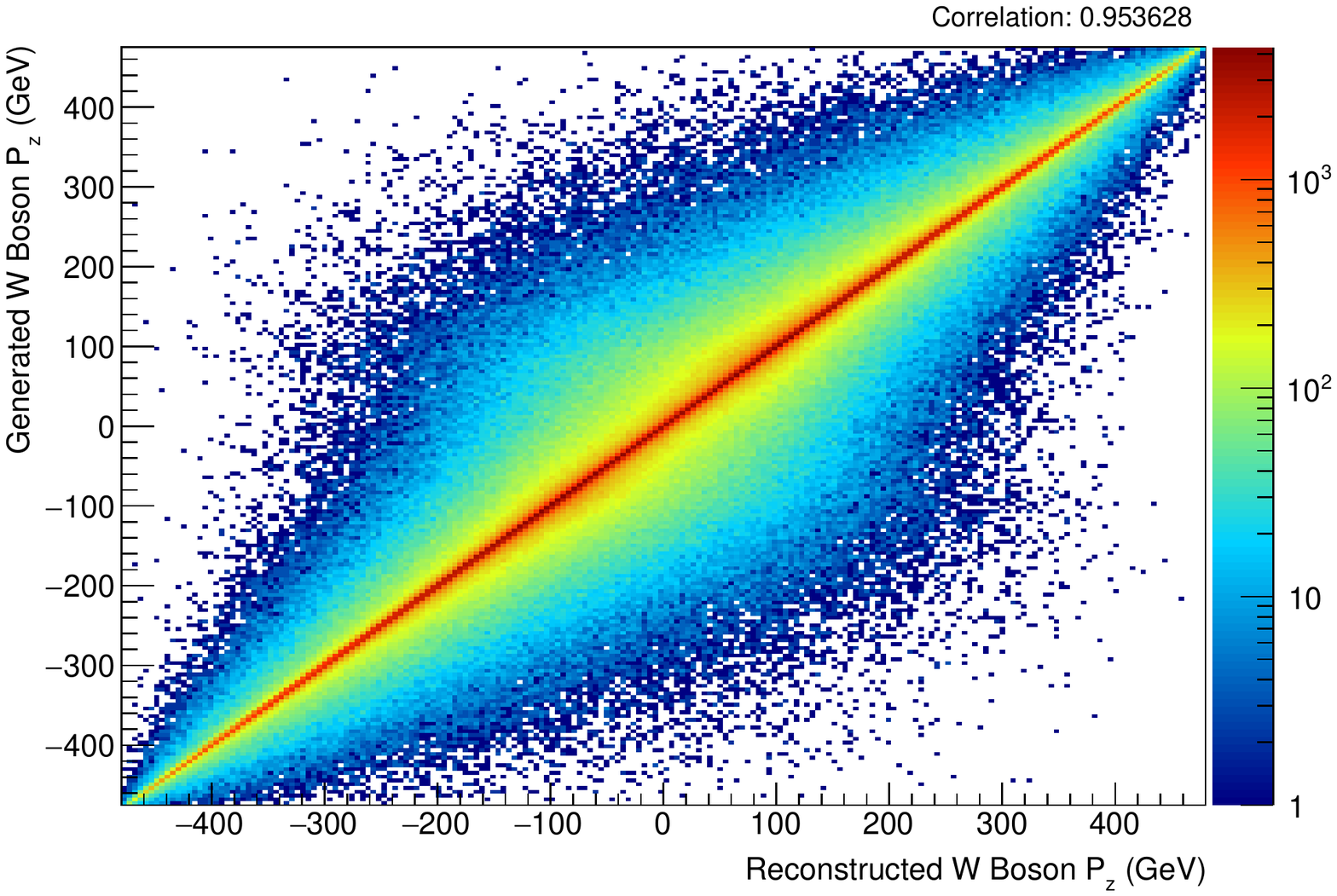} \\
\caption{Correlation plots between the generated and reconstructed transverse (left) and $z$-axis (right) momentum components of the $W^\pm$ bosons. These plots were produced with Sample A.}
\label{fig:random2}
\end{center}
\end{figure*}

The results of both methods are statistically compatible, with no real solution found for 12\% of all events, one real solution found with a frequency of 23\% and two real solutions per event found for the remaining 65\% of events. 
As a first step, in the case of an event with two possible real solutions, the solution to be considered in the reconstruction was selected randomly. (In section~\ref{sec:likelihood}, a likelihood method is introduced to select the most likely solution.) Figure~\ref{fig:random} shows the correlation plots between the generated \mbox{(anti-) neutrino} transverse and $z$-axis momentum components and their reconstructed values. The correlation between the generated and reconstructed neutrino kinematic variables is above 85\% percent, which indicates a successful reconstruction. It is also worth noting that the asymmetry seen between the $z$-axis momentum components of the neutrino and anti-neutrino is merely a consequence of the direction of the electron and positron beams upon the collision. In fact, because of the forward-backward asymmetry induced by the weak interaction, top quarks (and consequently neutrinos) are preferably emitted in the direction of the incoming electron, while anti-top quarks are preferably emitted in the direction of the incoming positron. As expected, the correlations between the generated and reconstructed transverse and $z$-axis momentum components of the $W^\pm$ bosons, shown in Figure~\ref{fig:random2}, are similar to the ones of the neutrinos. 

\begin{figure*}
\begin{center}
\includegraphics[width=8.cm]{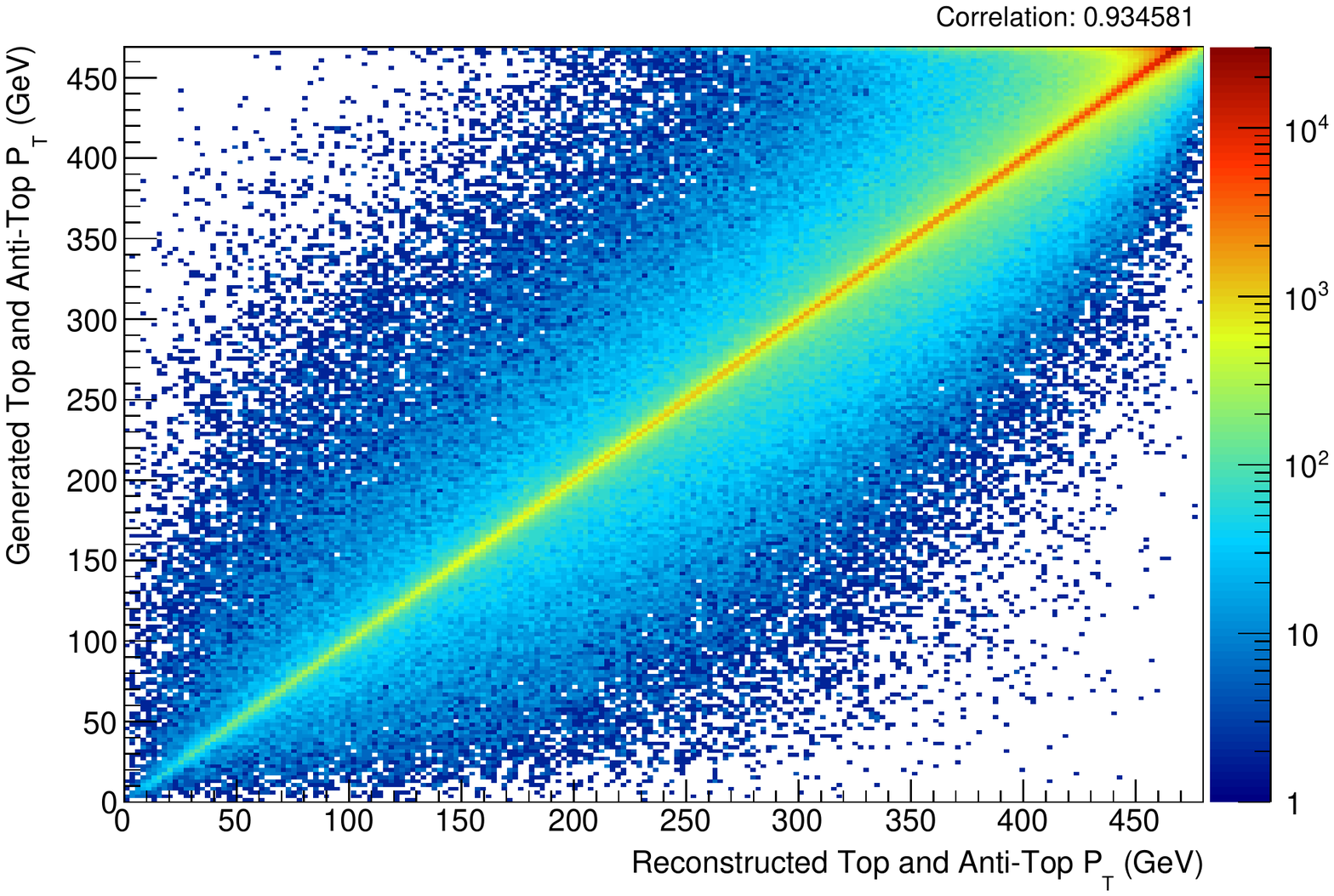}
\includegraphics[width=8.cm]{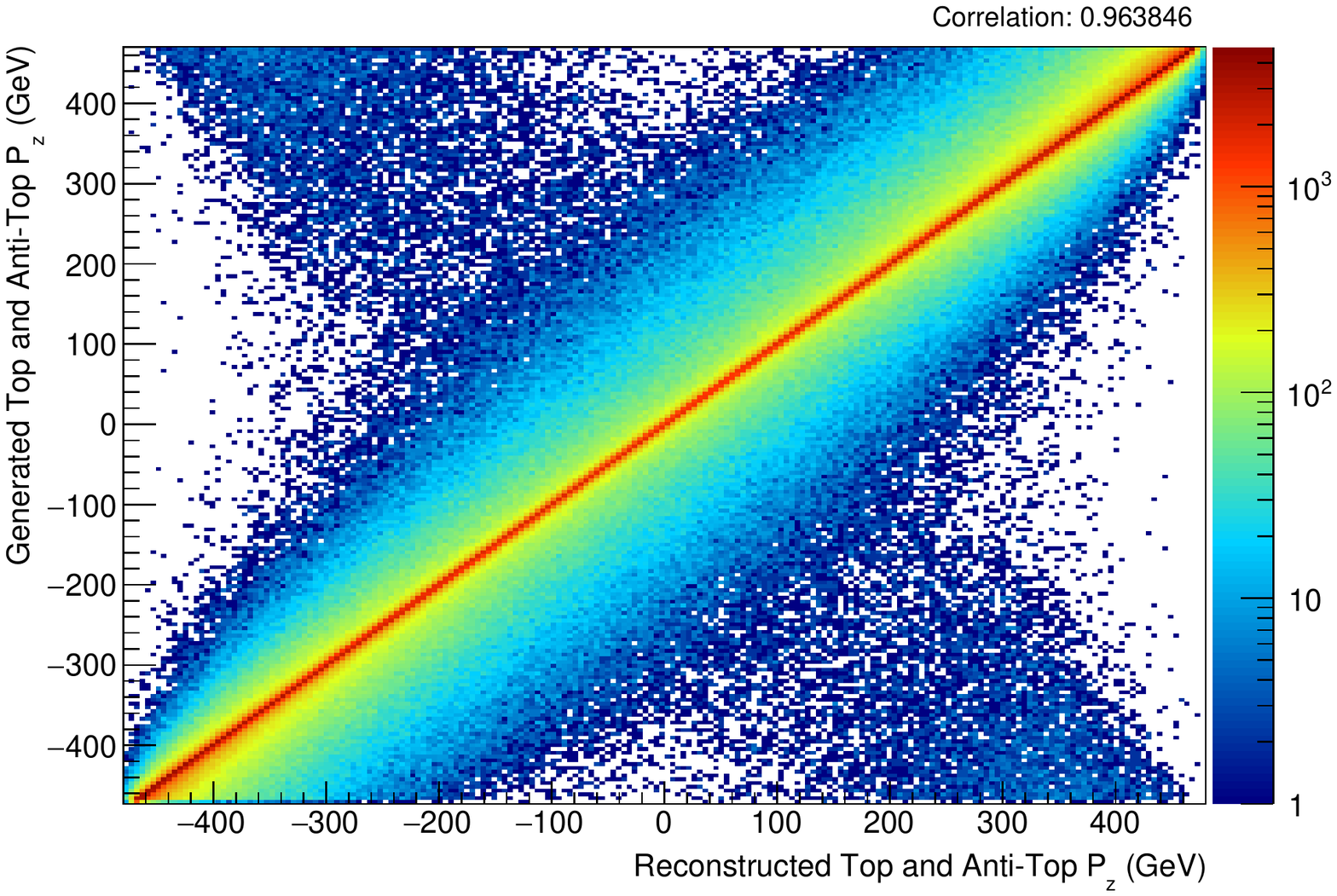} \\
\caption{Correlation plots between the generated and reconstructed transverse (left) and $z$-axis (right) momentum components of the top and anti-top quarks. These plots were produced with Sample A.}
\label{fig:random3}
\end{center}
\end{figure*}

The ultimate goal of this procedure is to fully reconstruct the momentum of the top and anti-top quarks. Since the charge of the $b$($\bar{b}$) jet is assumed to be unknown in the experimental analysis, one faces the problem of pairing the $b$($\bar{b}$) quark with the charged lepton resulting from the same $t$($\bar{t}$) decay. For each solution there are only two different pairing possibilities and the most likely pair is determined by means of a $\chi^2$ method. The $\chi^2$ variable employed in the code is defined as:
\begin{equation}
\chi^2 = \frac{(m_t^{\textrm{rec}}-m_t)^2}{\Gamma_t^2} + \frac{(m_{\bar{t}}^{\textrm{rec}}-m_{\bar{t}})^2}{\Gamma_{\bar{t}}^2} \, ,
\end{equation}
where $m_t^{\textrm{rec}}$ and $m_{\bar{t}}^{\textrm{rec}}$ are the reconstructed top and anti-top quark mass, respectively, with $m_t = m_{\bar t} = 173.2$~GeV and $\Gamma_t = \Gamma_{\bar t} = 1.42$~GeV~\cite{Khachatryan:2015hba}. The most likely pair candidate is determined by the lowest $\chi^2$ value. The correlation plots between the generated and reconstructed transverse and $z$-axis momentum components of the top and anti-top quarks are presented in Figure~\ref{fig:random3}. The $z$-axis momentum of the \mbox{(anti-) top} quarks on the right-hand side plot shows a residual anti-correlation. {The source of this anti-correlation was traced back to cases of wrong pairing between the $b$($\bar{b}$) quarks and the charged leptons. This problem can be addressed in the future by implementing more sophisticated statistical methods to establish the $b$-jet pairing.}

\section{Likelihood Method}
\label{sec:likelihood}

Since the system of kinematic equations may lead to two possible real solutions roughly 65\% of the time, one of the main challenges of the reconstruction procedure is to pick the right solution in these cases. In this study, a likelihood discriminant method was implemented in order to determine the most likely solution.
For each solution a likelihood variable, $\mathcal{L}$, is calculated as the product of several probability density functions (p.d.f., indicated with $P$ below). Three p.d.f.s were built from the top and anti-top quark kinematic variables, mass ($m_{t}$), transverse momentum ($p_{\textrm{T},t}$) and $z$-axis momentum component ($p_{\textrm{z},t}$),
\begin{equation}
\mathcal{L} = P(m_{t}) P(p_{\textrm{T},t}) P(p_{\textrm{z},t}) \, .
\label{eq:likelihood}
\end{equation}
The p.d.f.s in equation~(\ref{eq:likelihood}) were obtained by using an additional sample of one-million events generated with \verb!MadGraph5_aMC@NLO! at leading order, labelled as \mbox{Sample B}. Events in Sample B were generated assuming massless final state particles, with the exception of the bottom quarks, exactly as it was done for Sample A. Each solution for a given event in \mbox{Sample B} was assigned a ``good'' or ``bad'' label, based on the proximity between the reconstructed and generated \mbox{(anti-) neutrino}. The proximity criteria between the reconstructed and generated \mbox{(anti-) neutrino} is determined by means of a $\chi^2$ variable,
\begin{equation}
\chi^2 = \sum_{i=1}^{3}\frac{(p_{i}^{\nu,\textrm{rec}}-p_{i}^{\nu,\textrm{gen}})^2}{s} + \frac{(p_{i}^{\bar{\nu},\textrm{rec}}-p_{i}^{\bar{\nu},\textrm{gen}})^2}{s} \, ,
\end{equation}
where $p_{i}^{\nu,\textrm{rec}}$ and $p_{i}^{\bar{\nu},\textrm{rec}}$ correspond to the reconstructed neutrino and anti-neutrino momentum components, respectively. The generated neutrino and anti-neutrino momentum components are represented by $p_{i}^{\nu,\textrm{gen}}$ and $p_{i}^{\bar{\nu},\textrm{gen}}$, respectively.

\begin{figure*}
\begin{center}
\includegraphics[width=14.cm]{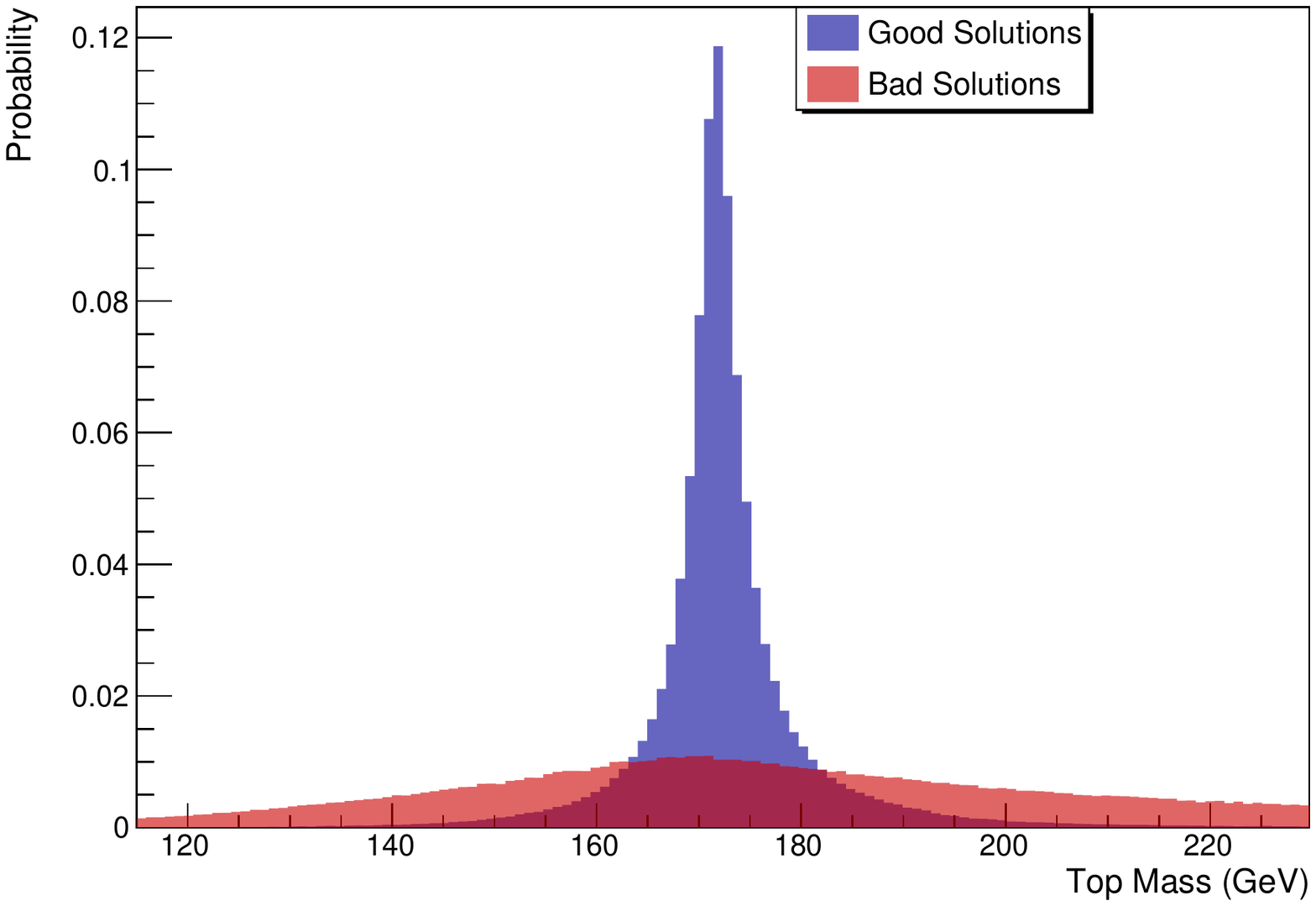}
\caption{Probability density functions (p.d.f.) of the \mbox{(anti-) top} quark mass. The blue shade represents the distribution for ``good'' solutions and the red shade represents the distribution for the ``bad'' solutions. These plots were produced with Sample B.}
\label{fig:pdf}
\end{center}
\end{figure*}

The \mbox{(anti-) top} quark mass p.d.f. distribution is shown in Figure~\ref{fig:pdf}, where the blue shaded histogram represents the distribution for ``good'' solutions and the red shaded histogram represents the distribution for the ``bad'' solutions. The difference in shape of the p.d.f.s for ``good'' and ``bad'' solutions provides significant discriminating power. The p.d.f.s were then used to calculate the likelihood variables for each solution of every event with two possible real solutions in Sample A. For each event solution in the Sample A, the likelihood of that solution being ``good'', $\mathcal{L}_G$, can be calculated as the product of the ``good'' solution p.d.f.s, using equation~(\ref{eq:likelihood}). In a similar fashion, the likelihood of that solution being ``bad'', $\mathcal{L}_B$, can also be calculated as the product of the ``bad'' solution p.d.f.s. Therefore, each event solution has a likelihood of being  ``good'' and ``bad''. For each event, the solution with higher likelihood ratio, ${\mathcal{L}_G}/{\mathcal{L}_B}$, is picked as the most likely candidate to be the ``good'' solution.

\begin{figure*}
\begin{center}
\includegraphics[width=8.cm]{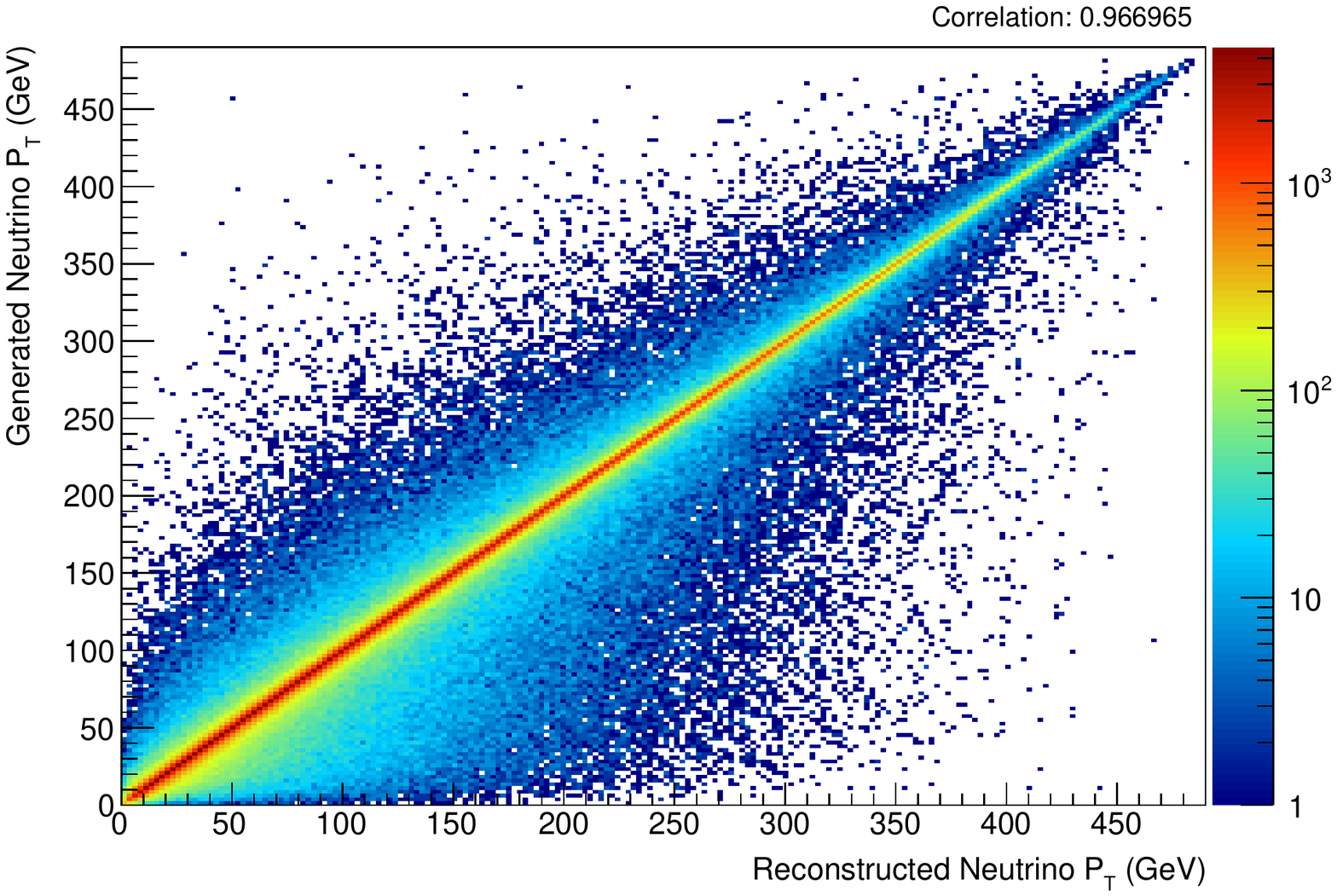}
\includegraphics[width=8.cm]{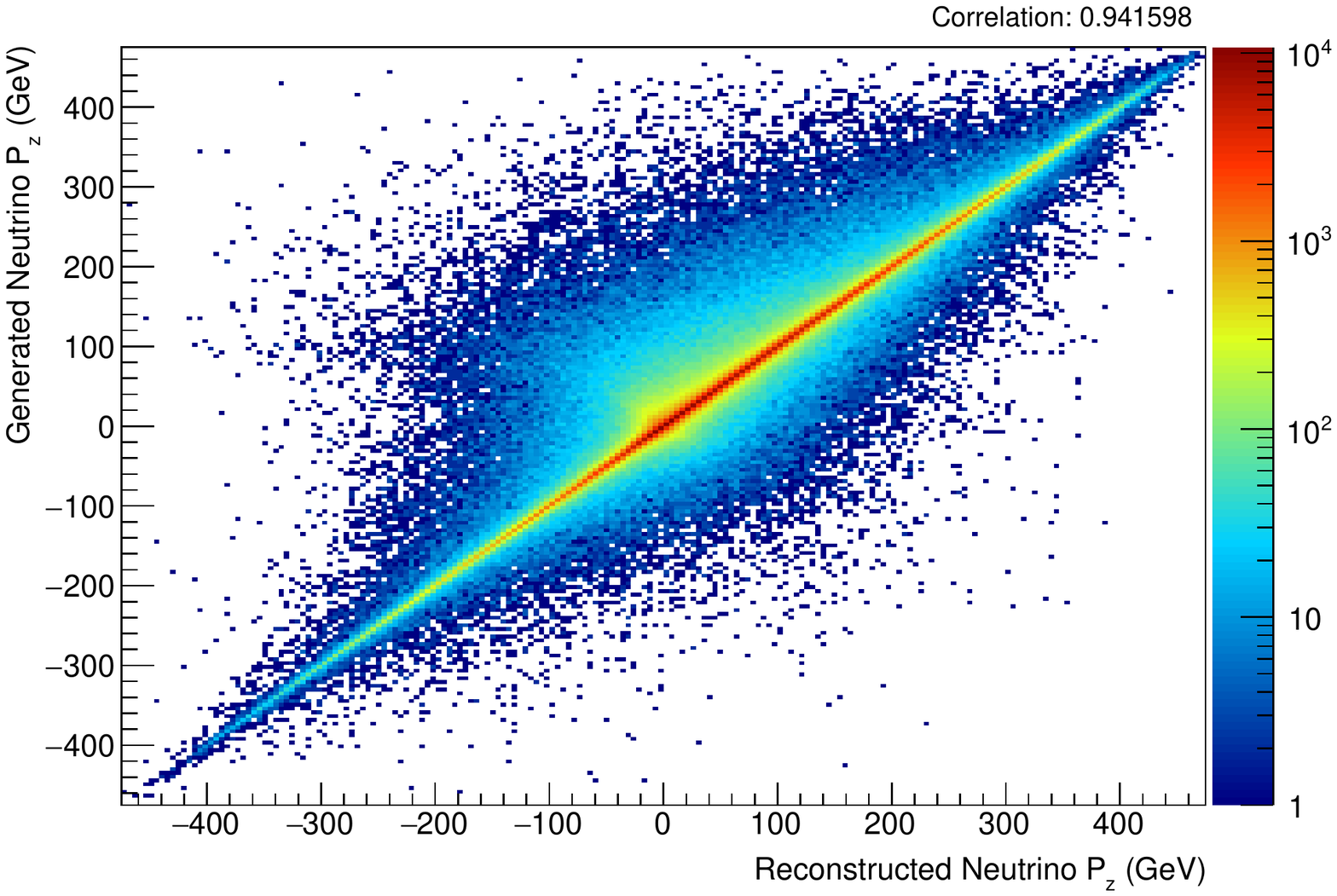} \\
\includegraphics[width=8.cm]{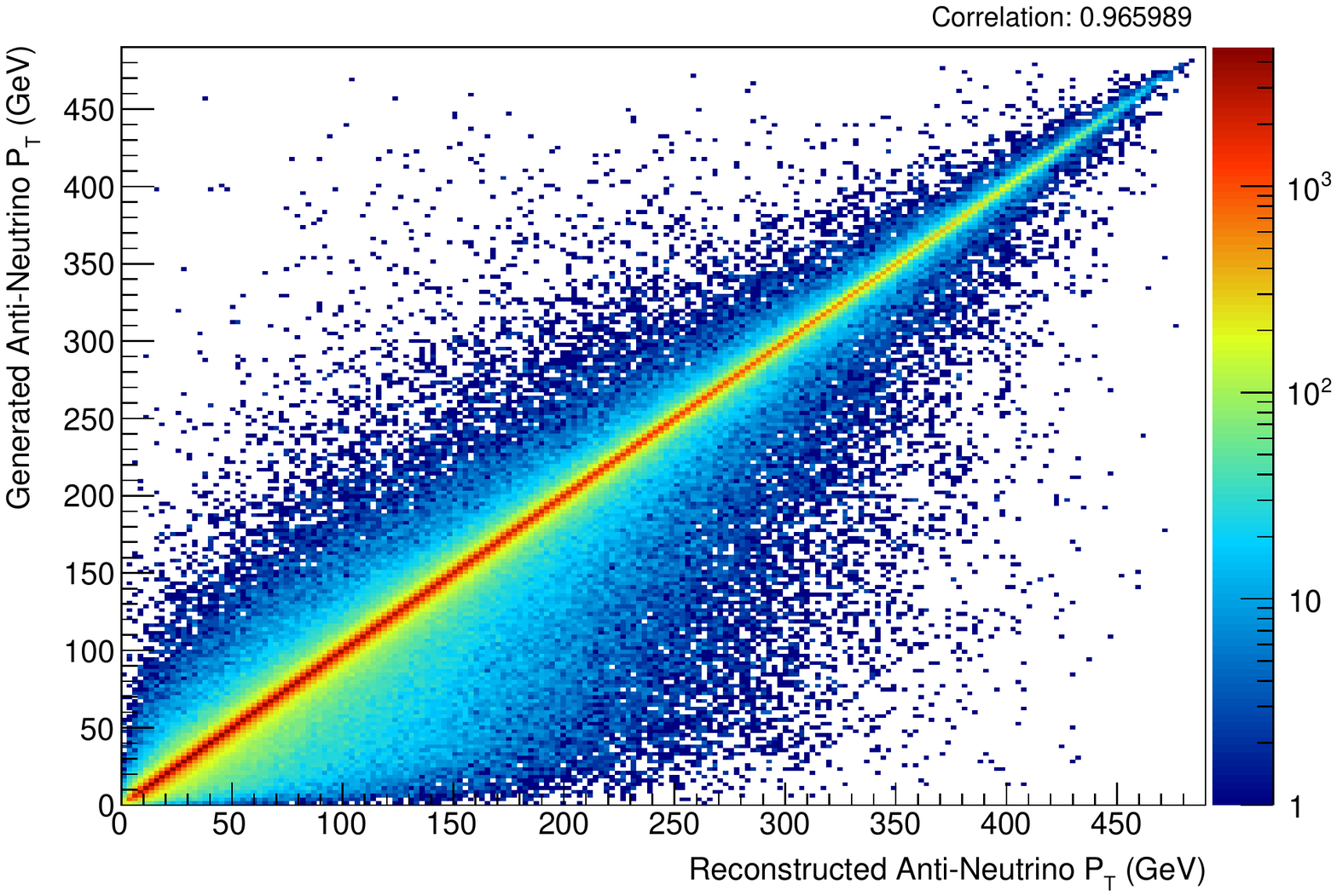}
\includegraphics[width=8.cm]{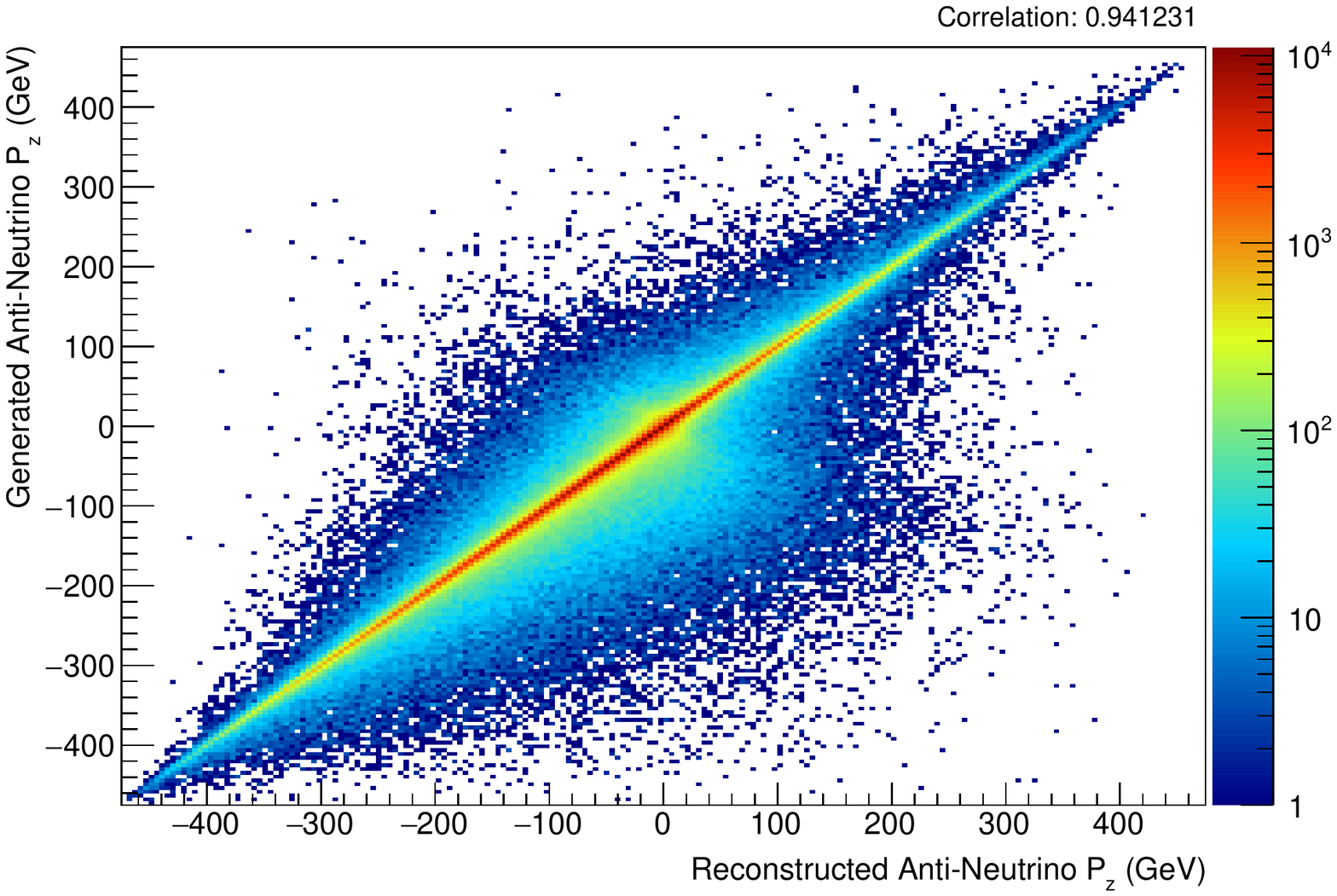}
\caption{Correlation plots between the generated and reconstructed transverse (left) and $z$-axis (right) momentum components of the neutrino (top) and anti-neutrino (bottom) after applying the likelihood method. These plots were produced with Sample A.}
\label{fig:neutrinosfinal}
\end{center}
\end{figure*}

\begin{figure*}
\begin{center}
\includegraphics[width=8.cm]{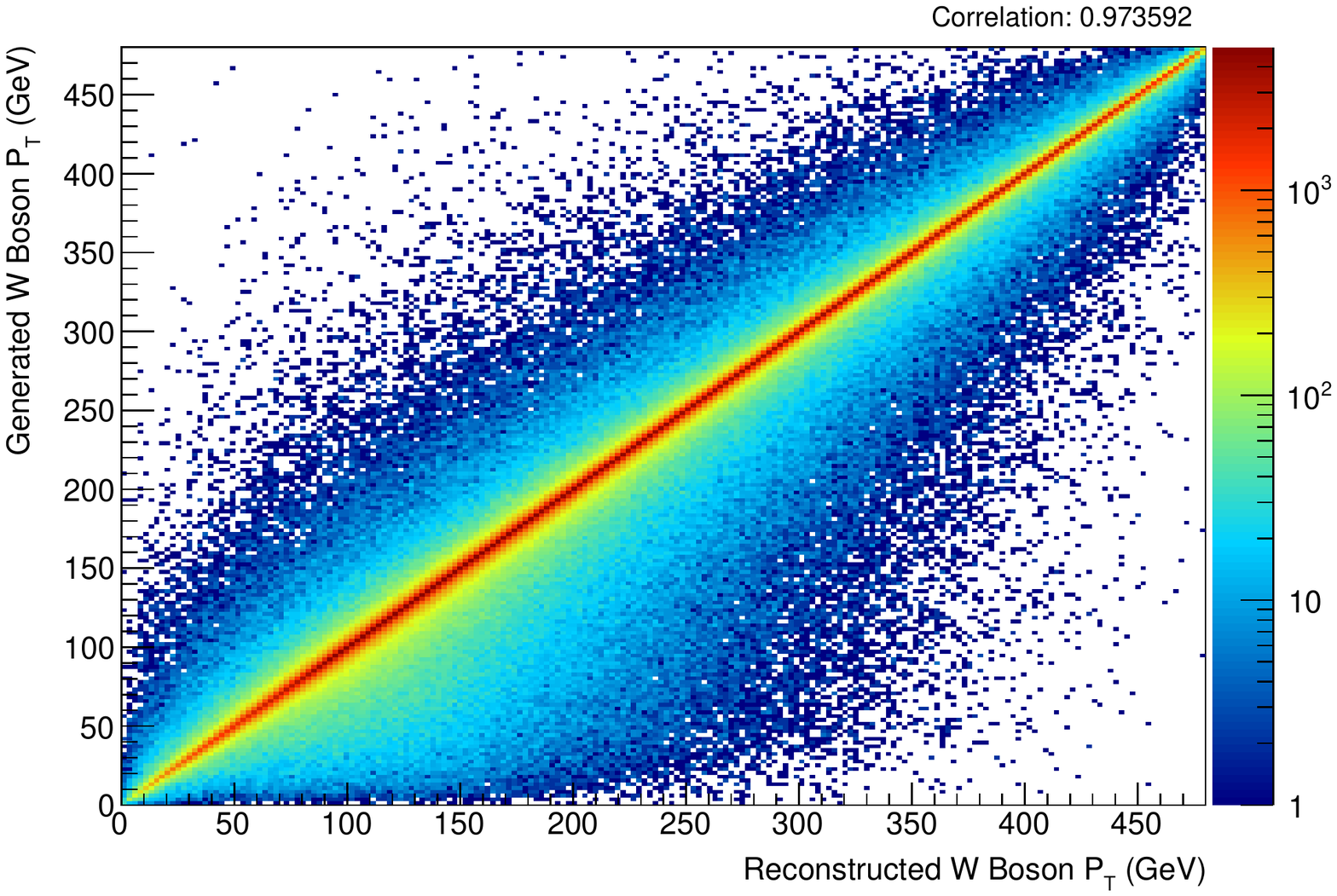}
\includegraphics[width=8.cm]{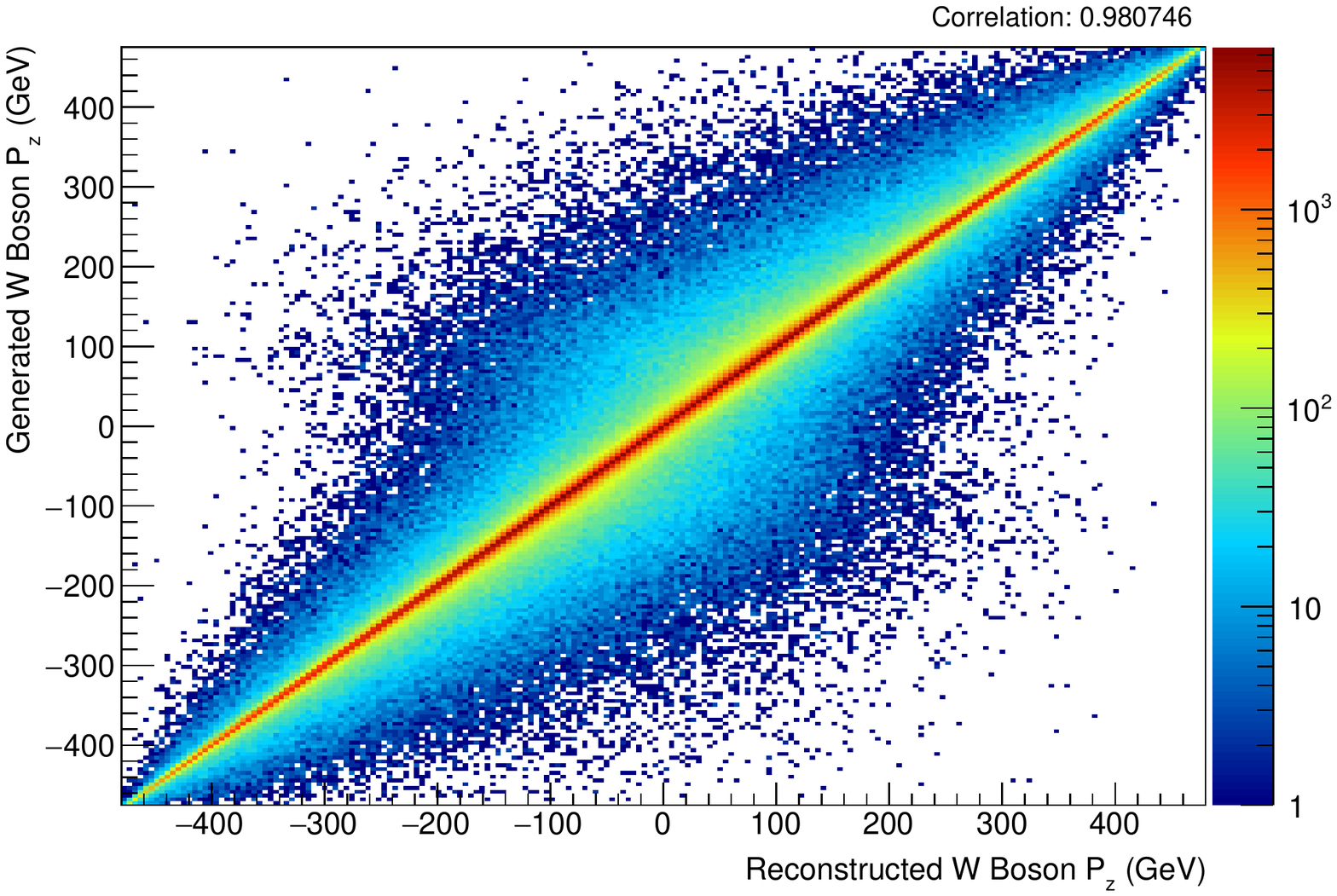} 
\caption{Correlation plots between the generated and reconstructed transverse (left) and $z$-axis (right) momentum components of the $W^\pm$ bosons after applying the likelihood method. These plots were produced with Sample A.}
\label{fig:wfinal}
\end{center}
\end{figure*}

\begin{figure*}
\begin{center}
\includegraphics[width=8.cm]{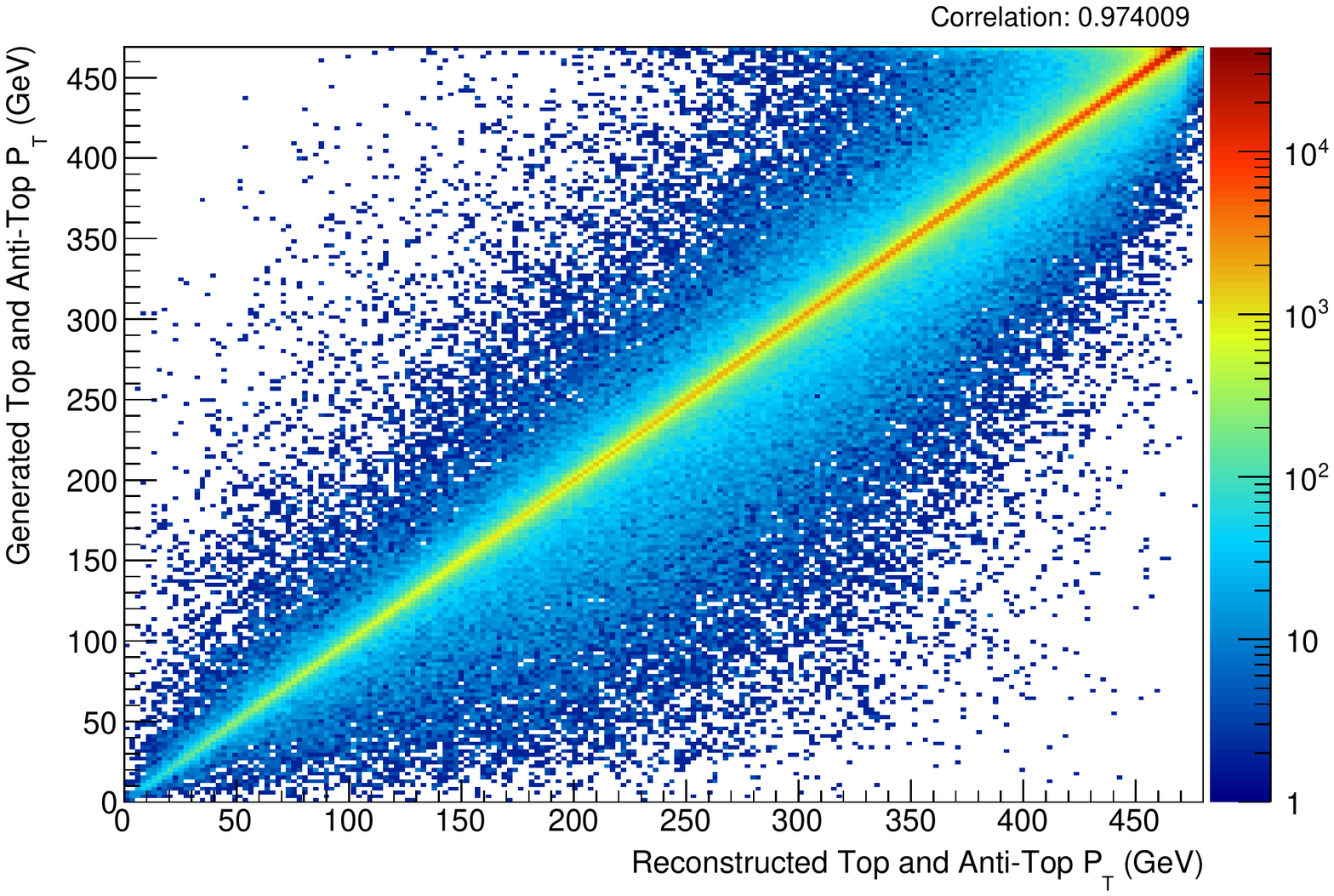}
\includegraphics[width=8.cm]{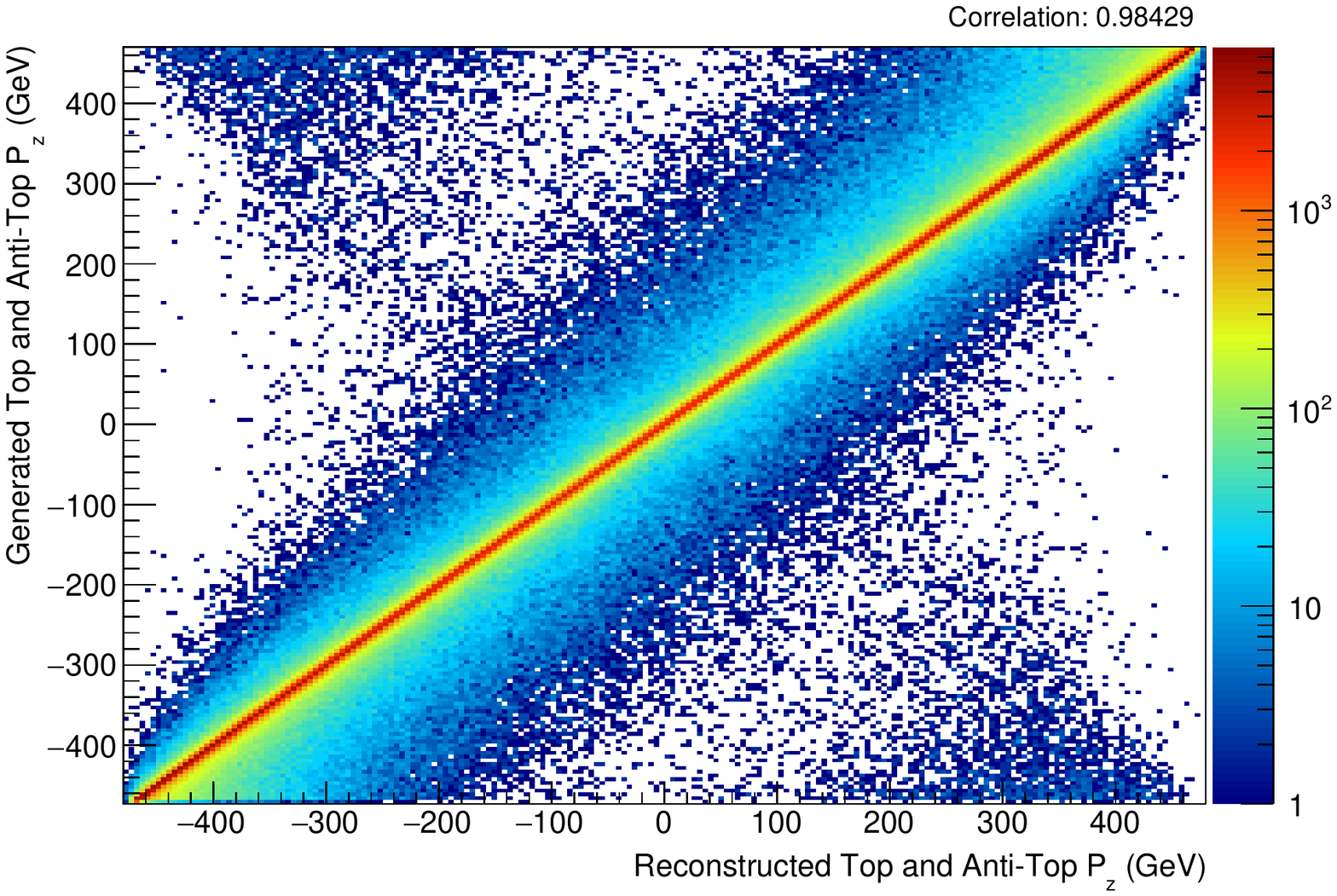} 
\caption{Correlation plots between the generated and reconstructed transverse (left) and $z$-axis (right) momentum components of the top and anti-top quarks after applying the likelihood method. These plots were produced with Sample A.}
\label{fig:topfinal}
\end{center}
\end{figure*}

Results obtained with Sample A are presented in Figures~\ref{fig:neutrinosfinal},~\ref{fig:wfinal}~and~\ref{fig:topfinal}. Figure~\ref{fig:neutrinosfinal} shows the correlation plots between the generated (anti-)~neutrino transverse and $z$-axis momentum components and their reconstructed values after applying the likelihood discriminant method. A correlation of about 95\% is obtained for these kinematic variables, a significant improvement when compared with the results of Figure~\ref{fig:random}. A clear improvement is also seen in the reconstruction of the $W^\pm$ bosons, shown in Figure~\ref{fig:wfinal}. Figure~\ref{fig:topfinal} shows the comparison between the generated (anti-)~top transverse and $z$-axis momentum components and their reconstructed values. A correlation above 95\% clearly indicates a successful reconstruction of the kinematic properties of this particle. 
The correlation between these reconstructed and generated kinematic variables can be further increased by applying a selection cut on the likelihood ratio variable.  Similar results can be achieved with other discriminant methods such as neural networks or multivariate analyses~\cite{Erdmann:2013rxa}. It should be stressed, however, that the effect of beam resolution, detector acceptance and selection cuts are expected to have an impact on the efficiency of the reconstruction procedure.

\section{Conclusions}
\label{sec:conclusions}

The goal of the present paper was to implement a method which allows for the reconstruction of the \mbox{(anti-) neutrino} momentum components in the dileptonic decays of a  top-quark pair at  future electron-positron colliders.

Two independent numerical methods were implemented for this purpose. The reconstruction code packages were thoroughly tested using generated samples of one-million electron-positron collision events at a center-of-mass energy of 1~TeV. The packages are publicly available and they can be downloaded from a repository \cite{link_to_package1,link_to_package2}.

In addition, a likelihood method was implemented to determine the most likely solution in each event.
If the likelihood method is applied, the correlation for the reconstructed \mbox{(anti-) neutrino}, the \mbox{(anti-) top} quarks and $W^\pm$ bosons is above 95\%. The effectiveness of the reconstruction package with and without the likelihood method can be evinced from the correlation plots found in Sections~\ref{sec:reconstruction} and \ref{sec:likelihood}.

The next step of this study will be to implement this reconstruction method in a dedicated analysis, in order to perform measurement estimations of top quark properties at a future electron-positron colliders. These estimations may include the study of top quark spin correlations, $W^\pm$ boson polarization in top quark decays and the top quark forward-backward asymmetry. This will require the simulation of a general purpose detector, the implementation of an event selection, and a detailed study of the different sources of systematic uncertainties. \\

\section*{Acknowledgments}
The authors would like to thank the Center for Theoretical Physics of the Physics Department at the New York City College of Technology, for providing computing power from their High-Performance Computing Cluster. This work was funded by PSC-CUNY Awards 61085-00~49 and 61151-00~49, and by FCT,
Lisboa 2020, Compete 2020, Portugal 2020, FEDER through project
POCI/01-0145-FEDER-029147. The work of A.F. was supported in part by the National Science Foundation under Grant No. PHY-1417354. The work of H.C. was supported in part by the Department of Energy under Grant No. DE-SC0019027. The authors would like to thank N. Castro for useful discussions and for reading the manuscript.


\end{document}